# Crowding and Queuing in Entrance Scenarios: Influence of Corridor Width in Front of Bottlenecks

**Juliane Adrian[1], Maik Boltes[1], Stefan Holl[1], Anna Sieben[2], Armin Seyfried[1,3]**
[1]Institute for Advanced Simulation, IAS-7: Civil Safety Research, Forschungszentrum Jülich
Jülich, Germany
j.adrian@fz-juelich.de; m.boltes@fz-juelich.de; st.holl@fz-juelich.de
[2]Chair of Social Theory and Social Psychology, Ruhr-Universität Bochum,
Bochum, Germany
Anna.sieben@rub.de
[3]School of Architecture and Civil Engineering, University of Wuppertal,
Wuppertal, Germany
a.seyfried@fz-juelich.de

***Abstract*** - In this paper, we present results of an entrance experiment investigating the effect of the corridor width in front of a bottleneck on the density. The idea is based on a previous study suggesting that a guiding system in front of an entrance can reduce pushing of the waiting people and thus the density at the entrance. In our study we aim to find out to what extend the corridor width has an impact on crowding or queuing behavior and with that on the density. The results of the presented experiment suggest that the transition takes place between a corridor width of 1.2 m and 2.3 m. The total duration of each experimental run is not significantly influenced by the corridor width but by the width of the entrance itself, the number of participants and partly by the motivation. In general, the density in front of the gate as well as the area of high density is increased by widening the corridor and by intensifying the motivation of the participants. However, the results also suggest that also the number of participants significantly influences the occurrence of pushing and the level of density.

***Keywords***: crowd management, entrance scenario, queuing, crowding, experiments

## 1. Introduction

Understanding the behavior of crowds in entrance scenarios is important to draw up or to adapt safety regulations for buildings and events. This understanding is based on observations, field studies and experiments as well as simulations. In a recently performed experiment we studied the transition between crowding and queuing behavior in entrance scenarios depending on the width of the corridor located in front of a bottleneck. The objective was to identify a relationship between density in the crowd and width of corridor, as well as the justness of the entrance procedure.

In a previous study, two entry situations were investigated using different types of spatial barrier structures positioned in front of a bottleneck [1]. In the first scenario, no barriers were used to guide the participants in front of the entrance. In this case, the participants were able to position themselves freely before the gates were opened. This resulted in a semicircle arrangement of the participants. In the second situation, barriers were used forming a corridor perpendicular to the entrance forcing the participants to arrange themselves in loose lanes before the gates were opened. After opening the gates, the participants in the semicircle setup showed crowding behavior. The crowd contracted as soon as the signal for entering was given. In case of the corridor setup, the participants formed loose lanes indicating the formation of an ordering structure, rather showing a queuing than a pushing behavior. On the one hand, the comparison between the distances from the entrance with the times in which the entrance is reached indicates that the semicircle setup is more just than the corridor setup. This is basically a consequence of the high densities in the semicircle setup which make it impossible to overtake persons in front. On the other hand, the high density makes the semicircle setup less safe, decreases the participants' well-being and leads to the perception of less justness. These results let to the question whether there is a critical corridor width limiting queuing behavior and stimulating pushing behavior instead. In order to investigate this question, we present density analysis and a comparison between waiting time versus distance to target for different corridor width and different motivation.

## 2. Methods

The presented experiments were carried out at the University of Wuppertal, Germany in January 2018. For each study, between 20 and 75 students were recruited as participants. The participants had to imagine a situation in which they want to enter a concert of their favorite artist. Each group had to perform two runs. In the first run, the motivation was higher than in the second run. This was communicated as follows: In the first run they had to imagine that an undisturbed view of the stage is only guaranteed for the first persons to enter. In the second run, the motivation was decreased by the announcement that all persons will have an undisturbed view of the stage, but nevertheless, the participants were told to enter quickly because they want to be close to the stage.

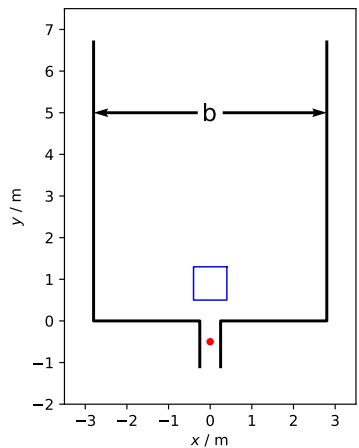

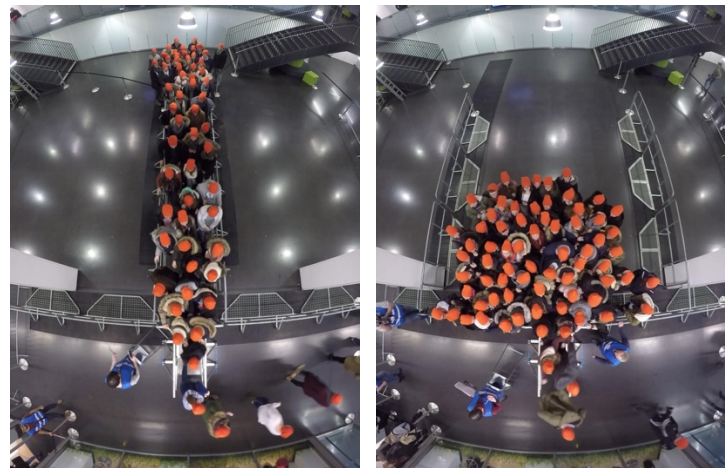

Figure 2: Snapshot for $t = 5$ s of Run 110 (left) and Run 030 (right).

Figure 1: Sketch of the experimental setup. The corridor width b is varied between 1.2 m and 5.6 m. Target is the entrance gate at $x = 0$ m, $y = -0.5$ m (red dot) with a width of 0.5 m. Blue rectangle: measurement area.

Table 1: Overview runs with different corridor width $b$, number of participants $N$ and motivation $h$ (0: high, -: low).

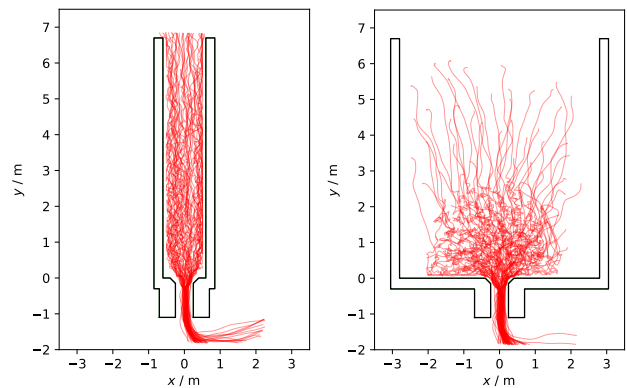

| Run | 110 | 120 | 230 | 240 | 270 | 280 | 050 | 060 | 030 | 040 |
|---|---|---|---|---|---|---|---|---|---|---|
| b | 1.2 m | | 2.3 m | | 3.4 m | | 4.5 m | | 5.6 m | |
| N | 63 | | 42 | | 67 | | 42 | | 75 | |
| h | 0 | - | 0 | - | 0 | - | 0 | - | 0 | - |

Figure 3: Trajectories of Run 110 with $b = 1.2$ m, $N = 65$ (left) and Run 030 with $b = 5.6$ m, $N = 75$ (right).

The experimental setup included an entrance gate with a width of 0.5 m and a corridor leading straight to the gate (see Figure 1). For different runs, the width of the corridor $b$ was varied between $b = 1.2$ m and $b = 5.6$ m to analyze the transition between low and high state. The number of persons passing the bottleneck has not been limited additionally. An overview over the corridor width and the number of participants is listed in Table 1 for a representative set of runs.

As an example, Figure 2 shows a screenshot for time $t = 5$ s after the start signal for two runs, one with the smallest corridor width $b = 1.2$ m (Run 110) and one with the largest corridor width $b = 5.6$ m (Run 030). It is clearly seen that the narrow corridor forces the participants to queue with a maximum of two to three persons side by side. The line of sight of all participants, and with that the preferred direction of movement, is oriented along the y-axis towards the target. Therefore, it is expected that most of the pressure is oriented in the same direction and the effective area is generally the area in front. In contrast, the participants in Run 030 are not forced into any formation because of the wide corridor. They arrange themselves in a semicircle in front of the entrance. This leads to a preferred direction of movement with a radial orientation.

Trajectories of individual participants were determined by automatic extraction from video recordings [2]. Figure 3 displays the trajectories corresponding to Runs 110 and 030. As expected from the snapshots in Figure 2, the trajectories of Run 110 show characteristics of lane formation whereas the trajectories of Run 030 show crowding characteristics after a contraction phase.

## 3. Results

Data analysis, such as density, velocity and flow measurements, were performed according to [3] using JuPedSim [4,5]. The results presented in the following concentrate on a representative set of runs, listed in Table 1.

### 3.1. Voronoi Density

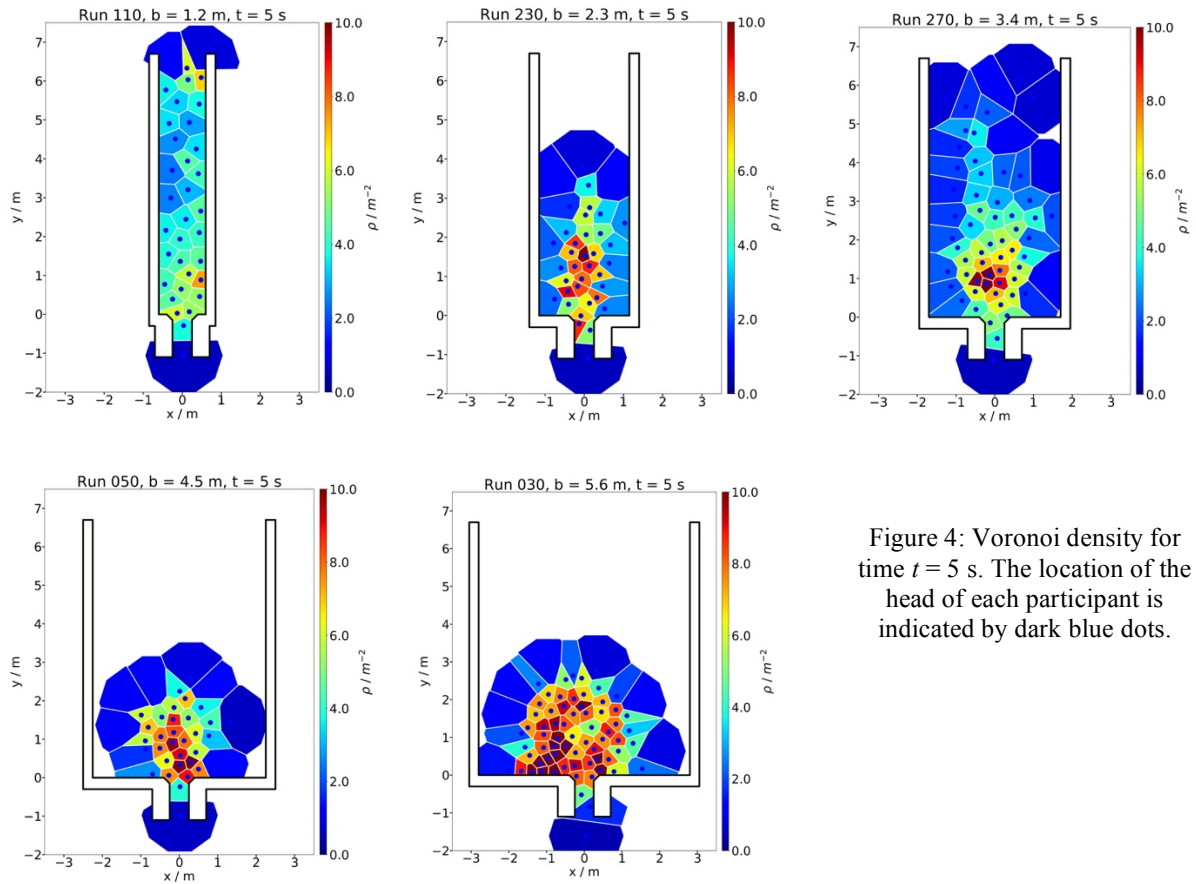

Figure 4: Voronoi density for time $t = 5$ s. The location of the head of each participant is indicated by dark blue dots.

Figure 4 shows the Voronoi cells and the location of each participant for $t = 5$ s after the start signal. At the chosen time, the contraction has already taken place. The color of the polygons indicates the density which is given by the reciprocal of the size of the individual Voronoi cell.

As seen in Figure 2, the narrow corridor of 1.2 m in Run 110 forces the participants into a rather ordered queuing arrangement. The densities within 3 m distance to the target are around 5 m$^{-2}$. In Runs 230 ($b = 2.3$ m) and 270 ($b = 3.4$ m), the arrangement of the participants in short distance to the entrance is rather v-shaped than a queue. Highest densities of partly more than 10 m$^{-2}$ are found within a limited area in ca. 1-2 m distance to the target. For both, Run 050 ($b = 4.5$ m) and Run 030 ($b = 5.6$ m), the wide corridor favors a semicircle arrangement in front of the entrance gate. This leads to densities of partly more than 10 m$^{-2}$ within 2 m distance to the target. In case of $b = 5.6$ m, these high densities are found in a larger area which is extended towards the left and right of the entrance gate. The participants form a semicircle clearly indicating crowding behavior.

### 3.2. Density time-series

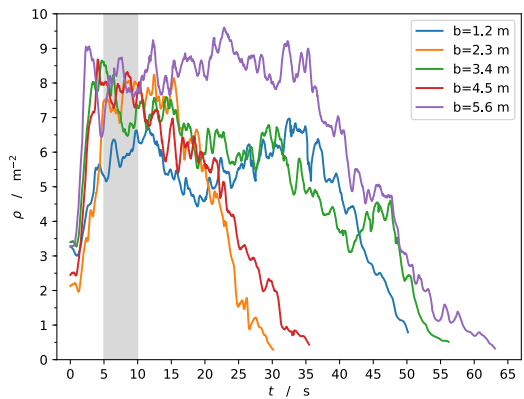 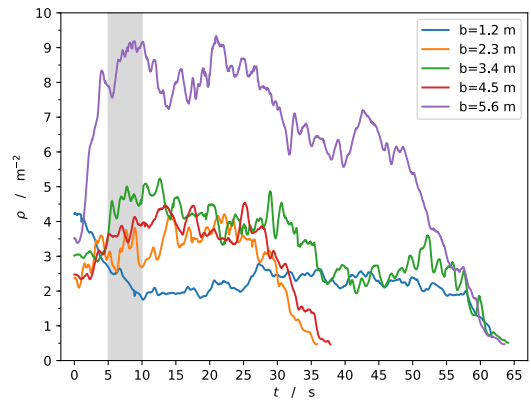

Figure 5: Time-series of density within the measurement area; high motivation.

Figure 6: Time-series of density within the measurement area; lower motivation.

The development of the density within the measurement area over time is shown in Figure 5 for runs with high motivation and in Figure 6 for runs with lower motivation. The size of the measurement area is 0.8 m by 0.8 m and its closest border is located in 0.5 m distance from the entrance of the gate (see Figure 1). The dimension and location of the measurement area is chosen so that it covers the area with highest density for all runs.

In case of high motivation, the density increases within ca. 5 s from an initial density up to a local density maximum indicating a contraction phase. This local density maximum is lowest for the smallest corridor width of 1.2 m. For the widest corridor ($b = 5.6$ m), the density stays at one level between 8 m$^{-2}$ and 9 m$^{-2}$ for about 40 s. The densities of the runs with $b = 4.5$ m, $b = 3.4$ m and $b = 2.3$ m reach a similar local density maximum as the run with $b = 5.6$ m. However, the density only shortly remains at one level and decreases afterwards. In comparison to $b = 4.5$ m and $b = 2.3$ m, the density time-series for $b = 3.4$ m does not decrease as rapidly. It reaches two more local maxima before decreasing to zero. For $b = 1.2$ m, the density increase takes longer than for the other runs indicating the absence of a rapid contraction. It reaches a maximum of 6.5 m$^{-2}$ after ca. 15 s. Afterwards, it decreases to a minimum of 5 m$^{-2}$ at $t = 20$ s and increases to a second maximum at $t = 35$ s before it finally decreases.

As seen in Figure 6, the maximum density is lower in case of lower motivation. This is true for all runs except for the widest corridor ($b = 5.6$ m). Here, the maximum is similar to the run with high motivation. For the smallest corridor ($b = 1.2$ m), however, the density even decreases from an initial density of 4 m$^{-2}$ to a level of only ca. 2 m$^{-2}$. For all corridor width, the slope of the initial density increase (or decrease) is not as steep as in Figure 5 with the high motivation.

Multiple local maxima, as seen in Runs 110 and 270, indicate sequences of pushing and non-pushing. The slope of the density decrease at the end of the runs seems to be similar for all datasets. This indicates that the density time-series corresponding to the last persons is only controlled by the outflow through the entrance gate and not by the corridor width.

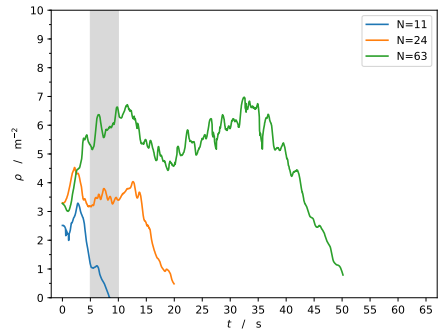

Figure 7: Time-series of density for varying number of participants N within the measurement area; high motivation, $b = 1.2$ m.

Another aspect that needs to be considered when comparing different runs, is the varying number of participants. In order to emphasize this assumption, Figure 7 displays density time-series measured in several runs with $b = 1.2$ m. The number of participants $N$ ranges from 11 to 63. It shows that not only the length of the run is dependent on the number of participants but also the maximum and mean density is.

Going back to Figure 5 and Figure 6, similarities between the time-series for $b = 2.3$ m and $b = 4.5$ m could be identified, such as maximum density, duration of the run and general curve shape of the time-series. This might be due to the fact that the same amount of people ($N = 42$) participated in these runs (see Table 1). In this case, the corridor width apparently has no big influence on the density. In contrast, in the runs with $b = 1.2$ m ($N = 63$) and $b = 3.4$ m ($N = 67$) nearly the same amount of people participated, but the density time-series show clear differences, especially for low motivation. Therefore, it is assumed that several factors, i.e., number of participants, corridor width and motivation, play a role. But, the influence of the individual factors cannot be quantified, based on the data at hand.

The density time-series for the largest width $b = 5.6$ m stands out. It shows the largest maximum density and the density is higher than for the other runs for a prolonged period of time. It is the run with the most participants ($N = 75$), but the difference in participants is only slightly higher than that for $b = 3.4$ m. Thus, it is more likely that those high densities stem from the difference in corridor width rather than from the difference in number of participants.

### 3.4. Influence of Corridor Width and Number of Participants on Mean Density

Since it is not suitable to define a steady state for all runs, we decided to use a 5 s interval from $t = 5$ s to $t = 10$ s to determine a mean density (gray area in Figure 5 and Figure 6). We assume that the contraction phase has already taken place after $t = 5$ s. Figure 8 displays the mean density in the 0.8 m by 0.8 m measurement area (see Figure 1) within this time-interval for all runs. In general, the mean density increases with increasing corridor width. For high motivation, it increases from ca. 6 m$^{-2}$ for Run 110 ($b = 1.2$ m) to ca. 8 m$^{-2}$ for $b = 4.5$ m (Run 050, 280) and only slightly less for $b = 5.6$ m (Runs 030, 150). Furthermore, there is a density gap of ca. 3-4 m$^{-2}$ between the runs with high motivation (h0) and the corresponding runs with low motivation (h-). This does not hold for the Runs 170, 070 and 190, which were conducted with high motivation. In Figure 8, the mean density of these runs fits to the group of low motivation rather than to the group with high motivation. The reason for these exceptions becomes clear in Figure 9. Here, the mean density is plotted versus the number of participants showing a clear relation. The number of participants in the mentioned runs is low in comparison to the other runs. Therefore, the density is also lower which fits to the findings of the density time-series (see Figure 7). According to Figure 9, the influence of the number of participants on density are not to be neglected. It seems that a certain amount of people is needed for the phases of contraction, congestion and release to fully build up.

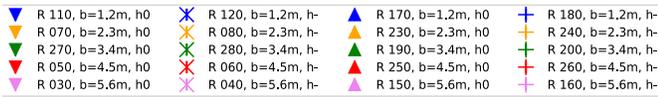
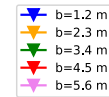
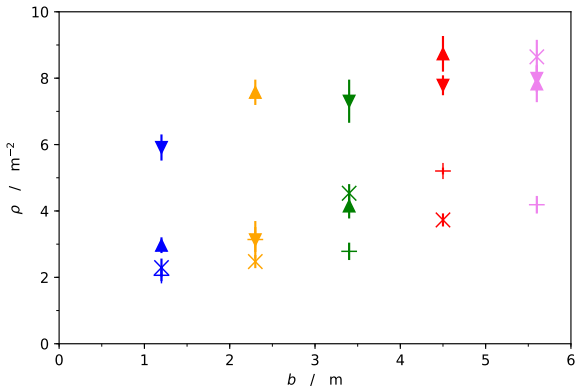
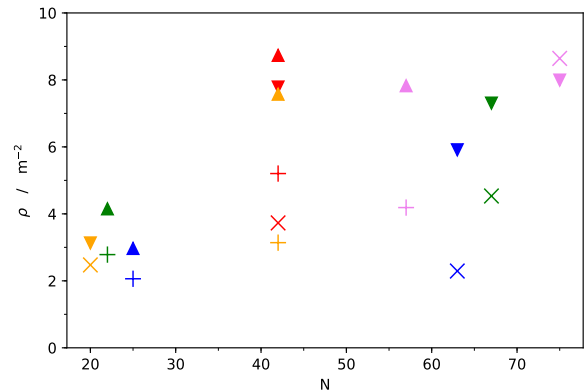

Figure 8: Mean density (determined from $t = 5$ s to $t = 10$ s) in the measurement area against corridor width including error bars. Triangles refer to high motivation, crosses and x to lower motivation. The color indicates the corridor width.

Figure 9: Mean density (determined from $t = 5$ s to $t = 10$ s) in the measurement area against number of participants. Triangles refer to high motivation, crosses and x to low motivation. The color indicates the corridor width (see Fig. 8).

### 3.5. Waiting Time and Distance to Target

Another characteristic of the entrance process which is helpful to distinguish between queuing and pushing behavior is the waiting time. Figure 10 compares the waiting time in dependence on the distance to the target for a narrow corridor of $b = 1.2$ m (left) with the corridor width of $b = 2.3$ m (center) and with the widest corridor of $b = 5.6$ m (right). For all three corridor width, the waiting times are displayed for both, a run with high motivation (top) and low motivation (bottom). For each participant, the blue dots represent the starting position in terms of linear distance to the target, which lies within the entrance gate (red dot in Figure 1), and the total time to reach the target. Please note that the trajectories were extracted within a limited area only. The maximum extension in y-direction was ca. 7.5 m. Therefore, the starting position of participants who gathered outside this area corresponds to the position where they stepped into the area of interest (see Runs 110 and 120).

During the first seconds after the start signal, two opposite behaviors or phases are observable: a rushing or contraction on the one hand and a waiting phase on the other hand. The contraction phase is most obvious in case of high motivation and can be observed for all shown corridor width: The individual lines run to the left nearly parallel to the x-axis. This indicates that the distance to the goal is reduced in a short period of time. Or with other words: The participants directly started moving in the direction of the target until all interspaces are filled and a certain (maximum) density is reached (see Figure 5 and Figure 6). In contrast, the waiting phase is indicated by lines parallel to the y-axis meaning that the distance to the target does not change significantly within a certain period of time. In this case persons start their movement only if the person in front has moved indicating queuing. Like in the transition from a stop wave to a go wave, it takes some time until the stop phase has resolved. This waiting phase is most obvious in case of the narrow corridor with low motivation (Run 120). In some runs, we observe a combination of both phases, e.g. in Run 040. However, in these cases the waiting phase is very short and the contraction phase is predominant.

After the initial phase, the participants find themselves in a congestion, i.e. they can only move forward when the persons in front are moving because all interspaces are filled. This intermediate phase is indicated by a high concentration of individual lines. The mean slope within the congestion phase represents the mean velocity of the individuals in direction of the target. The steeper the intermediate phase is, the lower is the velocity indicating the progress to the target, the gentler the slope is, the higher is the velocity.

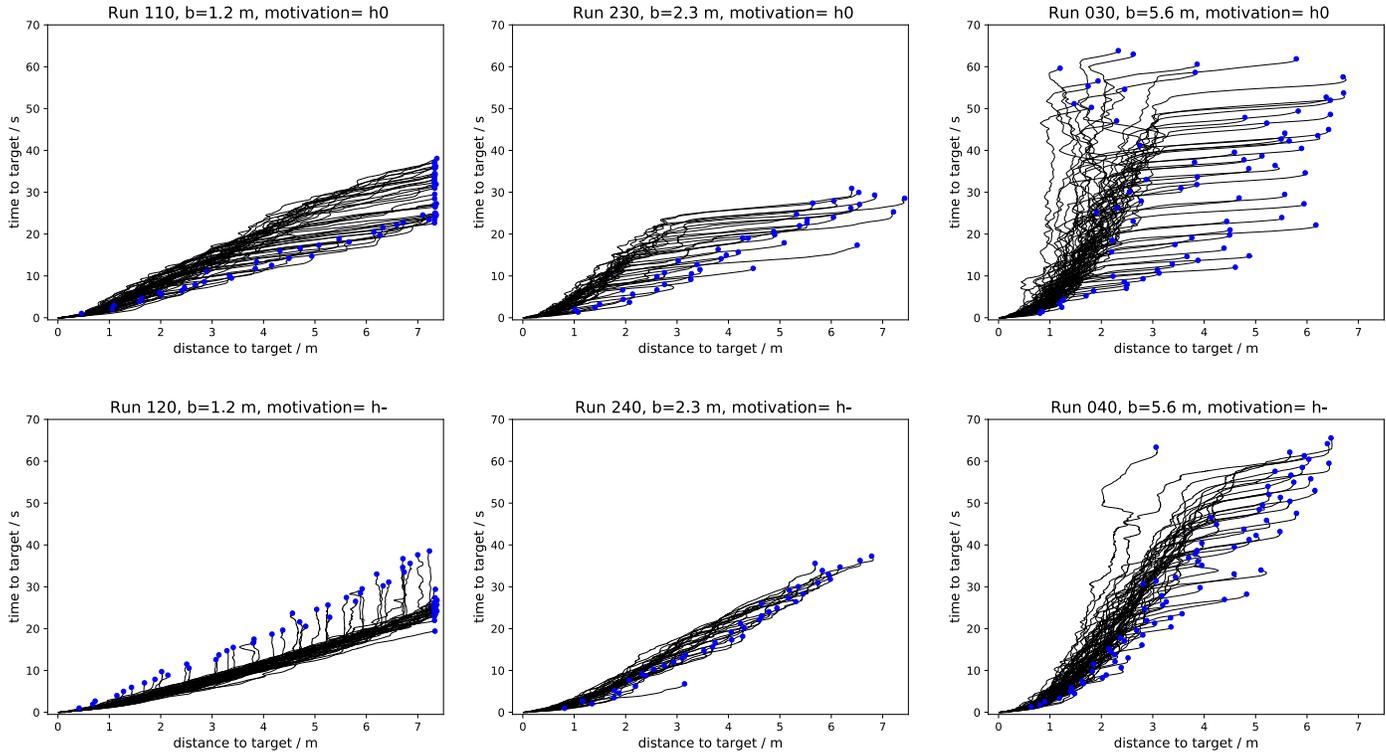

Figure 10: Waiting time versus linear distance to target. Blue dots indicate individual starting position and complete time to target. High motivation is indicated by h0, low motivation by h-, respectivielty. The target is located at $x = 0$ m, $y = -0.5$ m.

The final phase corresponds to the last 1 m in front of the target. The velocity is increased because the participants are less blocked by the congestion, i.e. there are less people between the release of the gate and the participant himself. This leads to a gentler slope in contrast to the intermediate phase meaning that more distance is made in a shorter period of time. This velocity increase can be seen in all presented runs.

In the following, differences or similarities between the different runs are examined to determine to what extend corridor width and motivation influences the waiting time in connection with the distance to the target entrance.

Apparently, high motivation leads to a pronounced contraction phase. In contrast, low motivation leads to a shorter contraction phase and facilitates a waiting phase which might be either additionally or instead. This indicates that the participants are willing to accept a higher density when being highly motivated.

For high motivation, the slope within the intermediate phase is steeper than that for the corresponding run with low motivation. In case of the narrowest corridor of $b = 1.2$ m, the slope of the intermediate phase is the shallowest whereas the slope is the steepest in case of the widest corridor (b=5.6 m).

Concerning the initial phase in Run 030 ($b = 5.6$ m), the following is observed: Some people are more active than others. This behavior shows in two ways: On the one hand, there are persons who are rushing in the very first seconds, but then enter a waiting phase meaning that they stand behind. On the other hand, the range of individual total runtime is quite broad for persons who started at a similar initial distance to the target. That means that some people are rushing quickly in the beginning to overtake some persons in front. Then they are slowed down when they reach the congestion.

The waiting time plots for $b = 2.3$ m (Runs 230 and 240) show more similarities with the ones corresponding to $b = 1.2$ m than to $b = 5.6$ m. Except that the waiting phase in the run with low motivation is not that pronounced.

## 4. Discussion and Conclusions

We investigated the influence of the corridor width in front of an entrance gate and the motivation on the behavior of participants in terms of arrangement, density and waiting time. Understanding this influence could help during the planning process and risk assessment for future building structures and for crowd management of events.

One of the main findings is that the corridor width and the motivation of the participants are important factors when determining whether people queue or start pushing. Furthermore, the density analysis revealed that the number of participants also has a significant influence on the maximum and mean density. This is true even for the narrowest corridor with a width of $b = 1.2$ m. Since the participants were recruited directly after their lesson, it was impossible to guarantee the same amount of participants for each run. In future experiments, the number should either be strictly controlled to ensure comparability or the number of participants should be used as additional parameter which is varied on a controlled basis.

The results of this study clearly reveal differences between wide and narrow corridors and also between different degrees of motivation. In case of the smallest corridor width of $b = 1.2$ m, the participants are forced into a queue with moderate densities. After the contraction phase, the participants have a rather large distance to the target, but they are constantly moving in target direction. Furthermore, the participants' well-being is maintained thanks to the comparably low densities. Overtaking is not necessary because of the constant movement. Therefore, social norms apply avoiding overtaking. In case of the widest corridor of $b = 5.6$ m, we observed a crowding behavior. The participants form a semicircle with high densities in front of the entrance gate but also to the left and right of the gate where people were pushed against the barriers. During the contraction phase overtaking was possible, but after that all interspaces were filled. In comparison to the narrow corridor, the individuals are closer to the target after the contraction, but then, during the intermediate phase, they spend more time in a congestion than with significant movement in target direction. The participants' well-being is reduced due to the high densities they are willing to accept.

The characteristics of the runs with intermediate corridor width, even for the second largest width of $b = 2.3$ m, indicate a pushing behavior rather than a queuing behavior, although the arrangement directly in front of the gate is mostly v-shaped instead of a semicircle. Generally, densities increase with increasing corridor width and increasing motivation.

Based on the presented results, it is assumed that the transition between pushing and queuing behavior takes place. At which width the transition takes place depends on the motivation, the number of people and on the width of the queuing system in front of the entrance. Therefore, we suggest to investigate more intermediate steps of corridor width in future experiments. Besides corridor width, motivation and the already mentioned number of participants, we suggest to consider heterogeneity of the group and to conduct more repetitions to ensure comparability and reproducibility and to increase the statistical reliability in order to derive the relation between the influence of the mentioned factors.